\documentclass[aps,prl,twocolumn,superscriptaddress]{revtex4}
\usepackage{graphicx,epsf}
\usepackage{amsmath}

\newcommand{\YRS}{YbRh$_2$Si$_2$}

\newcommand{\TK}{T_{\rm K}}

\begin{document}

\title{
Reply to Comment by V. R. Shaginyan {\em et al.} on\\
``Zeeman-Driven Lifshitz Transition: A Model for the
Experimentally Observed Fermi-Surface Reconstruction in \YRS''
}

\author{Andreas Hackl}
\affiliation{Department of Physics, California Institute of Technology, Pasadena, CA 91125, USA}
\author{Matthias Vojta}
\affiliation{Institut f\"ur Theoretische Physik, Technische Universit\"at Dresden,
01062 Dresden, Germany}

\date{\today}

\begin{abstract}
A reply to the comment by V. R. Shaginyan {\em et al.} [Phys. Rev. Lett. {\bf 107}, 279701 (2011),
arXiv:1206.5372] on our article [Phys. Rev. Lett. {\bf 106}, 137002 (2011), arXiv:1012.0303].
\end{abstract}

\maketitle


%
Our Letter \cite{L} proposed the scenario of a Zeeman-driven Lifshitz quantum phase
transition (QPT) to explain various thermodynamic and transport anomalies found in the
heavy-fermion metal \YRS, in particular their dependence on magnetic field and doping.
The Comment by Shaginyan {\em et al.} \cite{C} claims that
(a) our results, while qualitatively correct, are quantitatively incorrect because
we have ignored the temperature dependence of the quasiparticle density of states (DOS),
and that (b) our conclusions and predictions are artifacts of this approximation.
In the following we argue that these claims are either misleading or false.

(a) We start by pointing out that the purpose of \cite{L} was to discuss a scenario for \YRS\
which is {\em qualitatively} distinct from the popular idea of a Kondo-breakdown QPT. In
this discussion, quantitative details are less important, and naturally we made simplifying
assumptions in order to obtain explicit results.

It is true that we have neglected any temperature dependence of the DOS, the reason being
that we were interested in the low-temperature regime $T\ll\TK\approx 20$~K. In this
regime below the (effective) Fermi energy of the heavy-fermion system, one expects --
provided that a quasiparticle-based description is possible -- only weak $T^2$
Fermi-liquid corrections to the DOS \cite{FL} (whose quantitative calculation would require the
Fermi-liquid interaction functions as input which, however, are not known).
The expectation of weak low-temperature corrections to the DOS is {\em not} in
contradiction to the results of Refs.~\onlinecite{shim,ernst}, contrary to the claim in
\cite{C}, because those papers are concerned with the evolution of the DOS at much higher
temperatures 10~K\,$<T<$\,300~K.

We also note that a reliable {\em quantitative} calculation of the thermodynamic
properties at sub-Kelvin temperatures (which needs to involve both quasiparticle
interactions and collective effects) is beyond the scope of any available theoretical
method to date.

(b) As we made {\em qualitative} predictions in \cite{L}, and the authors of \cite{C} acknowledge our
considerations as being qualitatively correct, we feel that the claim in \cite{C} -- our
predictions being artifacts of our approximations -- makes the comment even internally
inconsistent.

In particular, the authors of \cite{C} claim that the power-law singularity (instead of a jump)
in the zero-temperature Hall coefficient upon passing the Lifshitz transition is such an
artifact. This claim is simply false. It is straightforward to calculate the Hall
coefficient from a Boltzmann treatment. This has been done in the literature for Lifshitz
transitions where Fermi pockets appear or disappear \cite{hackl09} and for other
Fermi-surface-topology-changing QPT \cite{norman}, with the consistent result that $R_H$
varies continuously in the weak-field limit (with a power-law piece near criticality).
Such a result is easy to rationalize: The contributions of a small pocket to the
components of the conductivity tensor vanish continuously as the pocket disappears
\cite{ziman}, which renders the $R_H$ evolution continuous in the presence of other
bands.

We note that such a continuous evolution at $T=0$ is not in contradiction to experiments
on \YRS\ \cite{friede10}, which provide data for $T>20$~mK: We have shown \cite{L} that
-- in the Lifshitz scenario -- the $R_H(B)$ curves at small finite $T$ have the form of a
smeared jump, with a width approximately linear in $T$ (above a crossover temperature set
by the critical field $B_c$), similar to the experimental data \cite{friede10}.

Finally, we speculate that some of the claims made in \cite{C} arise from the fact that the
authors think in terms of a ``fermion-condensation QPT'' \cite{shagi}
which, however, is rather different from the Lifshitz QPT alluded to in \cite{L}.

We conclude by mentioning that our Letter made a number of {\em novel} predictions, not
present in the literature on \YRS\ before. Most importantly:
(i) The low-field phase is a Fermi liquid at sufficiently low temperatures.
(ii) The smeared jump in $R_H$ at finite $T$ should not evolve into a sharp jump at $T=0$,
but remain continuous.
(iii) Carrier doping (instead of isoelectronic doping or pressure \cite{friede09}) should
allow to shift or even remove the $T^\ast$ line \cite{gegenwart07} from the phase diagram.
Notably, prediction (iii) has meanwhile led to new experiments \cite{geg} on Fe-doped
\YRS, which will contribute to settle on the nature of the QPT in \YRS.



\begin{thebibliography}{}

\bibitem{L}
A. Hackl and M. Vojta,
Phys. Rev. Lett. {\bf 106}, 137002 (2011).

\bibitem{C}
V. R. Shaginyan {\em et al.},
preceeding comment, Phys. Rev. Lett. {\bf 107}, 279701 (2011).

\bibitem{FL}
G. Baym and C. Pethick,
{\it Landau Fermi-Liquid Theory}, (Wiley, New York, 1991).

\bibitem{shim}
J. H. Shim, K. Haule, and G. Kotliar, Science {\bf 318}, 1615 (2007).

\bibitem{ernst}
S. Ernst {\em et al.}, Nature {\bf 474}, 362 (2011).

\bibitem{hackl09}
A. Hackl and S. Sachdev, Phys. Rev. B {\bf 79}, 235124 (2009).

\bibitem{norman}
Ya. B. Bazaliy, R. Ramazashvili, Q. Si, and M. R. Norman,
Phys. Rev. B {\bf 69}, 144423 (2004).

\bibitem{ziman}
J. M. Ziman, {\it Electrons and Phonons} (Oxford University Press, Oxford, 1960).

\bibitem{friede10}
S. Friedemann {\em et al.},
PNAS {\bf 107}, 14547 (2010).

\bibitem{shagi}
V. R. Shaginyan {\em et al.}, Phys. Rep. {\bf 492}, 31 (2010).

\bibitem{friede09}
S. Friedemann {\em et al.},
Nature Phys. {\bf 5}, 465 (2009).

\bibitem{gegenwart07}
P. Gegenwart {\em et al.},
Science {\bf 315}, 969 (2007).

\bibitem{geg}
P. Gegenwart, private communication.

\end{thebibliography}
\end{document}